\documentclass[aps,prl,onecolumn,superscriptaddress,groupedaddress]{revtex4}  
\usepackage{graphicx}  
\usepackage{color, soul}
\usepackage{dcolumn}   
\usepackage{bm}        
\usepackage{amssymb}   
\usepackage{amsmath}
\usepackage{physics}

\begin{document}


\newlength{\halfpagewidth}
\setlength{\halfpagewidth}{\linewidth}
\divide\halfpagewidth by 2
\newcommand{\leftsep}{%
\noindent\raisebox{4mm}[0ex][0ex]{%
\makebox[\halfpagewidth]{\hrulefill}\hbox{\vrule height 3pt}}%
}
\newcommand{\rightsep}{%
\noindent\hspace*{\halfpagewidth}%
\rlap{\raisebox{-3pt}[0ex][0ex]{\hbox{\vrule height 3pt}}}%
\makebox[\halfpagewidth]{\hrulefill} } 
    
\title{Photoconductivity calculations of bilayer graphene from first principles and deformation-potential approach}
    
\author{Yijun~Ge}
\affiliation{Mechanical and Aerospace Engineering Department, University of California, Los Angeles, CA 90095}        

\author{Timothy~S.~Fisher}
\affiliation{Mechanical and Aerospace Engineering Department, University of California, Los Angeles, CA 90095}

\date{\today}

\begin{abstract}
We report first-principles calculations of electron-phonon coupling in bilayer graphene and the corresponding contribution to carrier scattering. At the phonon $\Gamma$ point, electrons with energies less than 200 meV are scattered predominantly by LA$^\prime$ and TA$^\prime$ modes while higher-energy electron scattering is dominated by optical phonon modes. Based on a two-temperature model, heat transfer from electrons with an initial temperature of 2000 K to the lattice (phonons) with an initial temperature of 300 K is computed, and in the overall relaxation process, most of this energy scatters into K-point phonon optical modes due to their strong coupling with electrons and their high energies. A Drude model is used to calculate photoconductivity for bilayer graphene with different doping levels. Good agreement with prior experimental trends for both the real and imaginary components of photoconductivity confirms the model's applicability. The effects of doping levels and electron-phonon scattering on photoconductiviy are analyzed. We also extract acoustic and optical deformation potentials from average scattering rates obtained from density functional theory (DFT) calculations and compare associated photoconductivity calculations with DFT results. The comparison indicates that momentum-dependent electron-phonon scattering potentials are required to provide accurate predictions.
\end{abstract}
\setulcolor{blue}
\maketitle
\twocolumngrid 
\section{\label{sec:level1}I. INTRODUCTION}
Bernal stacked bilayer graphene exhibits a significantly different band structure compared to single-layer graphene. The conduction and valence bands split into two subbands with a separation of 0.4 eV between the two conduction subbands edges, and the linearity near the Dirac point is broken, accompanied by a reduction in the Fermi velocity. By breaking the inversion symmetry of the two layers, a band gap is induced and can be tuned through chemical doping \cite{yu2011} or gating \cite{zhang2009direct}, suggesting potential applications in transistors, optoelectronics and photonics \cite{bonaccorso2010graphene,bao2012graphene}. The performance of some of these devices depends strongly on the efficiency of carrier photoexcitation and subsequent hot carrier relaxation processes \cite{tielrooij2013photoexcitation}. The purpose of this work is to develop a model for the coupling of electrons and phonons that can predict electron relaxation processes due to phonon interactions from first principles. A secondary objective is to compare these predictions to the commonly used deformation potential scattering approach.

Photoexcited carriers thermalize rapidly over a timescale of 100 to 200 fs and thus can be described by an effective electron temperature $T_e$. The hot carriers thereafter then scatter with phonons and transfer energy to the lattice. This process lasts for tens of picoseconds until the system returns to thermal equilibrium. Optical pump terahertz probe spectroscopy \cite{george2008ultrafast,kar2018ultrafast} and angle-resolved photoelectron spectroscopy \cite{ohta2007interlayer,tanaka2013investigation} are the most often used methods for measuring electron-phonon coupling and carrier transport properties. The pump-probe method directly relates the measured optical transmission with and without photoexcitation to the photo-induced conductivity change in the material under study. The transition from negative to positive differential transmission has been observed in graphene sheets by changing electrostatic gating and the Fermi level \cite{shi2014controlling, frenzel2014semiconducting}. The transition is explained as an interplay between Drude weight and scattering rates. Photoexcitation generates more electron-hole pairs in intrinsic graphene but only modifies the Fermi distribution in doped graphene and therefore increases absorption in the former case while inducing more scattering in the latter case. However, only phenomenological scattering rates have been used in prior studies, leaving details of the scattering processes unresolved \cite{shi2014controlling,frenzel2014semiconducting,heyman2015carrier}. 

Prior theoretical work on bilayer graphene's electrical conductivity has focused on short-range scattering from impurities using tight-binding Hamiltonians \cite{adam2008boltzmann,ferreira2011unified}. Viljas et al.\ \cite{viljas2010electron} calculated electron-phonon heat transport based on a two-temperature model, with scattering rates obtained from an empirical deformation potential approach. Park et al.\ \cite{park2008electron} calculated electron-phonon coupling strengths from first principles; however, they focused on phonon lifetimes. Details of electron-phonon scattering mechanisms and their contributions to the electro-thermal transport properties of bilayer graphene are still not well understood. 

\begin{figure*}[htb]
\centering
\includegraphics [width=6 in]{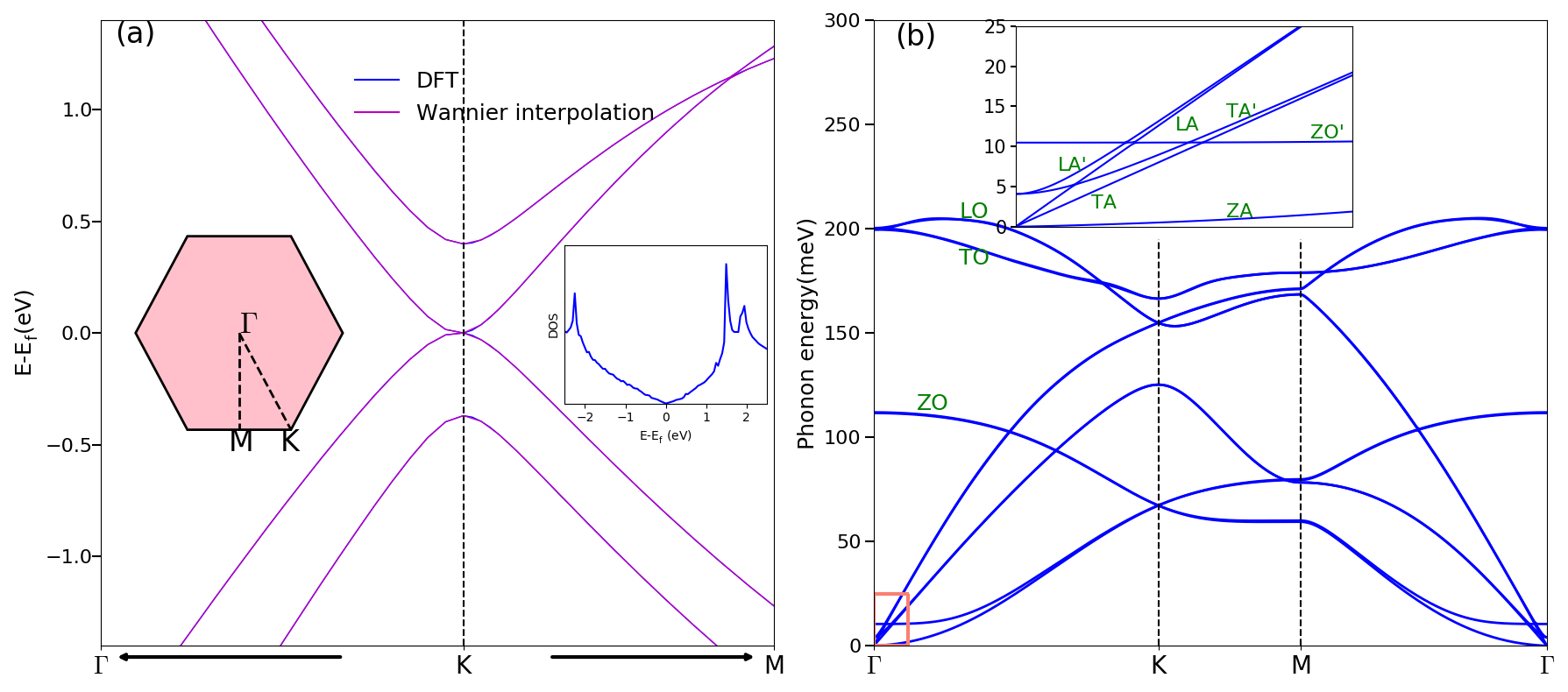}
\caption{Electron band structure and phonon dispersion of bilayer graphene. (a) Bilayer graphene electron band structure. (b) Bilayer graphene phonon dispersion. The inset of (b) shows a magnified view of the box at the bottom left corner. Two low energy branches which are absent from single-layer graphene, LA$^\prime$ and TA$^\prime$, are lifted and separated from the LA and TA modes near the $\Gamma$ point.}

\label{fig:F1}
\end{figure*}

In this paper, we adopt a first-principles approach to obtain electron-phonon coupling constants and scattering rates from which effective acoustic and optical deformation potentials are extracted. A two-temperature heat transfer model is developed to predict the electron and lattice temperatures at different times, based on which we calculate photoconductivity using a Drude model. We explain the trends for real and imaginary parts of photoconductivity by analyzing the effects of doping levels and electron-phonon scattering, and further compare photoconductivity from DFT and the deformation-potential approach with previous experimental data.

\section{\label{sec:level2}II. Electron phonon interactions, deformation- potential and heat transfer}

The development of density functional perturbation theory(DFPT) \cite{baroni2001phonons}  has made first-principles calculations of electron-phonon coupling feasible in recent years. The temperature-independent electron-phonon coupling matrix $g$ describes the coupling strength of the transition from an initial electron state $\ket{m,\textbf{k}}$ to a final state $\ket{n,\textbf{k}+\textbf{q}}$, and is calculated as \cite{lam1986self}: 

\begin{equation}
g_{mn\upsilon}(\textbf k,\textbf q) = \sqrt{\frac{\hbar}{2M\omega_{\textbf{q},\upsilon}}}\bra{n,\textbf{k}+\textbf{q}}\Delta V_{\textbf{q},\upsilon} \ket{m,\textbf{k}}
\label{eqng}
\end{equation}

\noindent
Here, $\ket{m,\textbf k}$ and $\ket{n,\textbf{k}+\textbf{q}} $ are wavefunctions for the initial and final Bloch states, $\Delta V_{\textbf{q},\upsilon} $ is the self-consistent potential change experienced by electrons due to interaction with a  phonon of wavevector $\textbf{q}$ in branch $\upsilon $, $M$ is the mass of carbon atoms, and $\omega_{\textbf{q},\upsilon}$ is the phonon frequency. 

After obtaining the electron-phonon coupling matrix, electron scattering rates are calculated based on Fermi's golden rule and the relaxation time approximation \cite{allen1987theory}:

\begin{equation}
\begin{split}
\tau_{m\textbf{k}}^{-1}&(T_e,T_{ph}) = \frac{2\pi}{\hbar} \sum_{n,\textbf{q},\upsilon}|g_{mn\upsilon}|^2 \\
\{&(f_{n,\textbf{k}+\textbf{q}}+n_{\textbf{q},\upsilon}) 
\delta(\epsilon_{m,\textbf{k}}+\hbar\omega_{\textbf{q},\upsilon}-\epsilon_{n,\textbf{k}+\textbf{q}}) \\
+&(1+n_{\textbf{q},\upsilon}-f_{n,\textbf{k}+\textbf{q}})\delta({\epsilon_{m,\textbf{k}}-\hbar\omega_{\textbf{q},\upsilon}-\epsilon_{n,\textbf{k}+\textbf{q}}})\}
\label{eqntau}
\end{split}
\end{equation}

\noindent
In Eq.\@ \ref{eqntau}, the first and second terms on the right side denote phonon absorption and emission. $f$ is the Fermi-Dirac distribution at the electron temperature $T_e$, and $n$ is the Bose-Einstein distribution at the phonon temperature $T_{ph}$. The summation extends over all electronic bands with a final state momentum $\textbf{k}+\textbf{q}$, phonon wavevectors and phonon branches. 

Similarly, the heat transfer rate per unit area from electrons to the lattice is computed as \cite{low2012cooling, vallabhaneni2016reliability}:

\begin{equation}
\begin{split}
Q&_{e-ph} = \frac{4\pi}{\hbar A} \sum_{\textbf{k},\textbf{q},m,n,\upsilon}\hbar\omega_{\textbf{q},\upsilon}|g_{m,n,\upsilon}(\textbf{k},\textbf{q})|^2 \\
\{&f_{n,\textbf{k}+\textbf{q}}(1-f_{m,\textbf{k}})(n_{\textbf{q},\upsilon}+1) 
-(1-f_{n,\textbf{k}+\textbf{q}})f_{m,\textbf{k}}n_{\textbf{q},\upsilon}\} \\
\delta&(\epsilon_{m,\textbf{k}}+\hbar\omega_{\textbf{q},\upsilon}-\epsilon_{n,\textbf{k}+\textbf{q}})
\label{eqnenergy}
\end{split}
\end{equation} 

\noindent
Here, $A$ is the unit cell surface area. The dynamic temperature evolution is governed by energy conservation as:
\begin{equation}
\begin{cases}
C_e(T_e)\frac{dT_e}{dt} = -Q_{e-ph}\\
C_{ph}(T_{ph})\frac{dT_{ph}}{dt} = Q_{e-ph}
\end{cases}
\label{eqnevol}
\end{equation}

\noindent
where $C_e$ and $C_{ph}$ are electron and phonon specific heats, respectively, and diffusive heat conduction has been neglected.

Even though first-principles calculations can be applied to any system in theory, the computations are extremely time-consuming. While Wannier function interpolation reduces computational cost greatly, the disentanglement procedure for complex material systems is quite involved, especially for interfaces consisting of dissimilar materials and having strong interactions.

For electron-phonon coupling calculations, the deformation-potential approach provides an alternative. Assuming that the perturbation of the potential felt by electrons is proportional to the change in unit cell volume, the deformation-potential approach is semiclassical and often obtained by fitting experimental mobility values. Here, we briefly discuss the derivation of the deformation potential for acoustic phonons. The development highlights the connection between the deformation potential and scattering rates, thus providing an explanation of how to extract the widely used deformation potential from our first-principles calculations.

The potential perturbation $\Delta V(\textbf{r})$ is related to the relative volume change $\Delta (\textbf{r})$ by \cite{bardeen1950deformation}: 

\begin{equation}
\Delta V(\textbf{r}) = E_1 \Delta (\textbf{r})
\label{eqndp}
\end{equation} 
where $E_1$ is the deformation potential, and $\textbf{r}$ denotes lattice sites. $\Delta(\textbf{r})$ is calculated in the long wavelength limit as:

\begin{equation}
\Delta(\textbf{r}) = \frac{\partial \textbf{u}(\textbf{r})}{\partial \textbf{r}}
\label{eqnderive}
\end{equation}

\noindent
Here, $\textbf{u}(\textbf{r})$ is the phonon displacement:

\begin{equation}
\textbf{u}(\textbf{r})=\sum_{\textbf{q}}{\textbf{e}_\textbf{q}}(A_{\textbf{q}}e^{i\textbf{q}\boldsymbol{\cdot}\textbf{r}}+A_{\textbf{q}}^{\ast}e^{-i\textbf{q}\boldsymbol{\cdot}\textbf{r}})
\label{eqndisp}
\end{equation}
where $\textbf{e}_\textbf{q}$ is the normalized phonon eigenvector, and $2|A_{\textbf{q}}|$ is the oscillation amplitude. Considering equipartition in which all phonons are excited and  $\hbar \omega_{\textbf{q}} \ll k_B T_L$, we can simplify the amplitude as  $\displaystyle |A_{\textbf{q}}^2| = \frac{k_B T}{2\rho A \omega_{\textbf{q}}^2}$ ($\rho$ is the mass density) by connecting the classical wave energy to the quantum harmonic oscillators' energy \cite{lundstrom2009fundamentals}. Combining Eqs.\@ \ref{eqndp}, \ref{eqnderive} and \ref{eqndisp}, we find:
\begin{equation}
|\Delta V|^2 = \frac{{E_1}^2k_B T_L}{2\rho A v_{\textbf{q}}^2}
\label{finalgkk}
\end{equation}
where $v_\textbf{q}$ is the acoustic phonon group velocity. The  transition probability $W_{\textbf{kk'}}$ from state $\ket{\textbf{k}}$ to state $\ket{\textbf{k}^\prime}$ is calculated as:

\begin{equation}
W_{\textbf{k,k}^\prime} = \frac{2\pi}{\hbar} |\Delta V|^2\delta(\epsilon_\textbf{k}\pm \hbar\omega_{\textbf{q}}-\epsilon_{\textbf{k}^\prime})\delta(\textbf{k}\pm \textbf{q}-\textbf{k}^\prime)
\label{transit}
\end{equation}

\noindent
The momentum relaxation rate then becomes:
\begin{equation}
\tau_{ADP}^{-1}(\epsilon_{\textbf{k}}) = \sum_{\textbf{k}^\prime}(1-\cos\theta(\textbf{k},\textbf{k}^\prime)) W_{\textbf{k},\textbf{k}^\prime}\frac{1-f(\epsilon_{\textbf{k}^\prime})}{1-f(\epsilon_{\textbf{k}})}
\label{relax}
\end{equation}
where $\cos\theta$ is the angle between $\textbf{k}$ and $\textbf{k}^\prime$. We assume that acoustic deformation potential scattering is elastic and therefore use the momentum relaxation rate to approximate the scattering rate. The subscript in $\tau_{ADP}^{-1}(\epsilon_{\textbf{k}})$ indicates that the scattering rate is derived based on the acoustic deformation potential.

Inserting Eqs.\@ \ref{finalgkk} and \ref{transit} into Eq.\@ \ref{relax} and considering valley and spin degeneracy, we obtain the scattering rate under the high-temperature and elastic-acoustic deformation approximations \cite{hwang2008acoustic}:

\begin{equation}
\tau_{ADP}^{-1}(\epsilon_{\textbf{k}}) = \frac{1}{4\hbar^3}\frac{\epsilon_{\textbf{k}}}{V_F^2}\frac{{E_1}^2}{\rho v_{\textbf{q}}^2}k_B T_L
\label{taufinal}
\end{equation}
where $V_F$ is the Fermi velocity. We assume an isotropic Fermi velocity $V_F = 6 \times 10^5 $ m/s in our calculations.

The average scattering rate is expressed as \cite{hwang2008acoustic}:

\begin{equation}
\langle\tau_{ADP}^{-1}\rangle = \frac{\int{d\epsilon D(\epsilon)\frac{1}{\tau(\epsilon)}(-\frac{\partial f}{\partial \epsilon})}}{\int{d\epsilon D(\epsilon)(-\frac{\partial f}{\partial \epsilon})}}
\label{avg}
\end{equation}
\noindent
where $D(\epsilon)$ is the density of states expressed as $\displaystyle \frac{2\epsilon}{\pi\hbar^2V_F^2}$. Eq.\@ \ref{avg} can be further reduced using Eq.\@ \ref{relax} as: 
\begin{equation}
\langle\tau_{ADP}^{-1}\rangle = \frac{E_1^2}{4 \rho v_{\textbf{q}}^2}\frac{k_B T_L}{\hbar^3V_F^2}\frac{\int{d\epsilon \epsilon^2 (-\frac{\partial f}{\partial \epsilon})}}{\int{d\epsilon \epsilon (-\frac{\partial f}{\partial \epsilon})}}
\label{redavg}
\end{equation}
By equating the average scattering rates from DFT and the deformation-potential framework, effective deformation potentials can be extracted. 

The optical deformation potential can be derived similarly with its scattering rate expressed as:
\begin{equation}
\tau_{ODP}^{-1}(\epsilon_{\textbf{k}}) = \frac{{D_o}^2}{4\rho \omega_o \hbar^2 {V_F}^2}[(\epsilon_\textbf{k}-\hbar \omega_o)(n_q+1)+(\epsilon_\textbf{k}+\hbar \omega_o) n_q]
\label{tauopt}
\end{equation}
where $\hbar \omega_o$ is the optical phonon frequency and $D_o$ is the optical deformation potential. The first term in the square bracket denotes the phonon emission, and only electrons with energies greater than $\hbar\omega_o$ are involved. 

\section{\label{sec:level3}III. PHOTOCONDUCTIVITY AND DRUDE MODEL}

\begin{figure}[htb]
\centering
\includegraphics [width=\columnwidth]{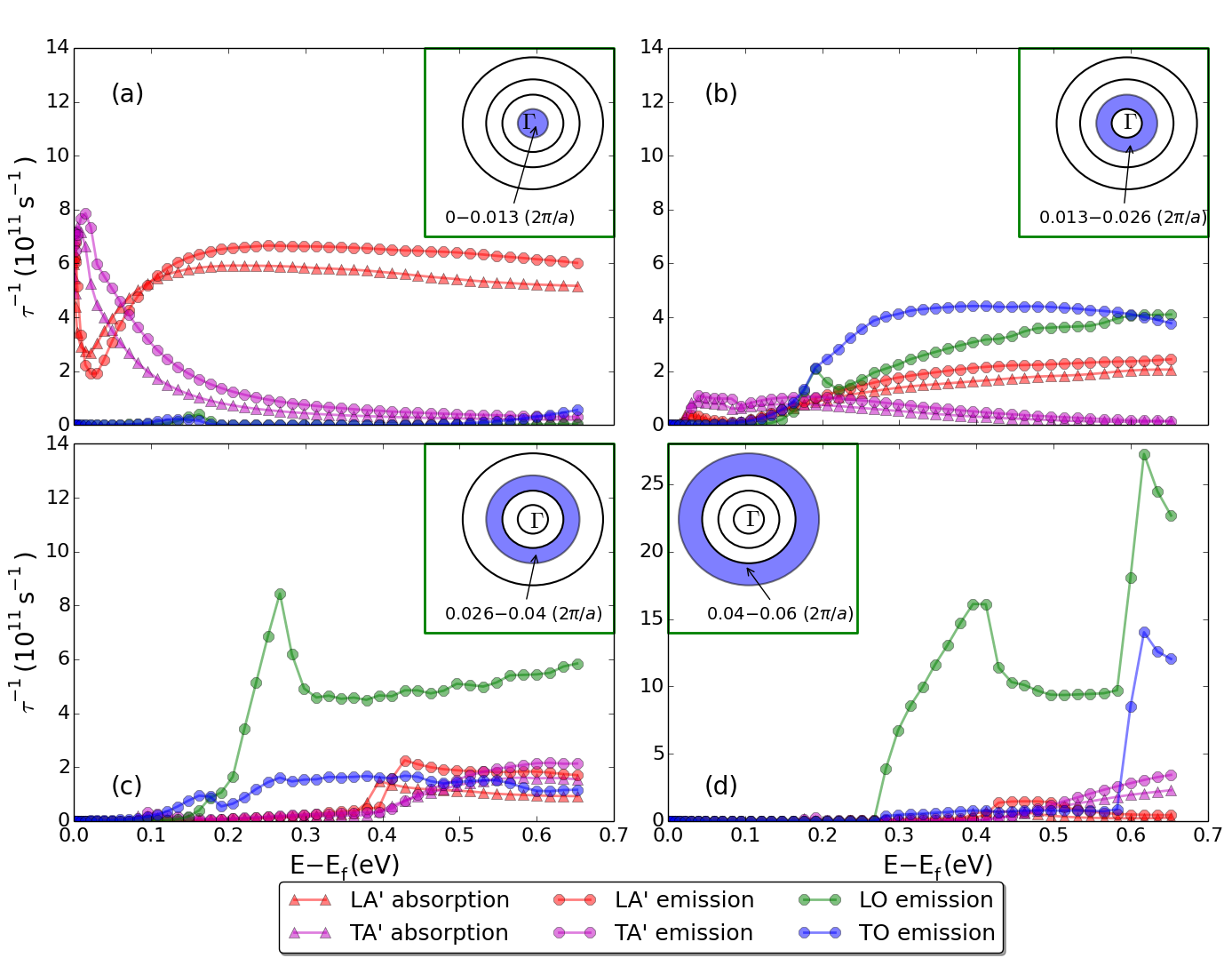}
\caption{Intrinsic bilayer graphene electron scattering rates near the phonon $\Gamma$ point at $T_e = 300$ K and $T_{ph}$ = 300 K. (a) Phonon wavevectors $\mathrm{0-0.013}\ 2\pi/a$. LA$^\prime$ and TA$^\prime$ modes dominate for electron energies less than 200 meV. (b) Phonon wavectors $\mathrm{0.013-0.026}\ 2\pi/a$. (c) Phonon wavevectors $\mathrm{0.026-0.04}\ 2\pi/a$. (d) Phonon wavevectors $\mathrm{0.04-0.06}\ 2\pi/a$. Optical modes only participate in phonon emission processes. 
}

\label{fig:F2}
\end{figure}

In pump-probe experiments, the variation of light transmission at a given frequency $\Delta T(\omega)$ is directly related to the photoconductivity $\Delta\sigma(\omega)$ via $\displaystyle \Delta\sigma(\omega)=-\frac{n_s+1}{Z_0}\Delta T(\omega)/T(\omega)$, where $T(\omega)$ is the transmission without photoexcitation, $\Delta T$ is the difference in transmission with and without photoexcitation, $n_s$ is the substrate's index of refraction, and $Z_0$ is the impedance of free space.

We use the Drude model for calculating photoconductivity. This model has proven to work well for bulk materials. Here, we consider bilayer graphene without defects throughout the paper; therefore the Drude model is a reasonable choice. According to this model, conductivity is expressed as:
\begin{equation}
\sigma(\omega)=\frac{ne^2}{m^{\ast}}\frac{1}{\tau^{-1}-i\omega}=ne\mu\frac{\tau^{-1}}{\tau^{-1}-i\omega}
\label{eqncond}
\end{equation}

\noindent
Here, $n$ is the carrier density, $e$ is the elementary charge, $m^\ast$ is the effective mass, and $\tau$ is the scattering rate.
We compare our simulations to experiments in which bilayer graphene with low defects was placed above the $\mathrm{SiO_2}$ substrate. Our current photoconductivity calculations do not consider surface optical phonon scattering as its contribution to the overall scattering rate is significantly smaller than intrinsic phonon scattering due to strong screening effects \cite{kar2018ultrafast}. These terms are related to electron mobility via $\displaystyle \mu=\frac{e\tau}{m^\ast}$, and $\mu$ can be calculated from:
\begin{equation}
\mu = \frac{eV_F^2}{2n}\int_{0}^{\infty}D(\epsilon)(-\frac{\partial f}{\partial \epsilon})\tau(\epsilon)d\epsilon
\label{eqnmu}
\end{equation}

\noindent
where $D(\epsilon)$ is the density of states. Inserting Eq.\@ \ref{eqnmu} into Eq.\@ \ref{eqncond} gives:
\begin{equation}
\sigma(\omega)=\frac{e^2V_F^2}{2}\int_{0}^{\infty}D(\epsilon)(-\frac{\partial f}{\partial \epsilon})\frac{1}{\tau^{-1}-i\omega} d\epsilon
\label{eqncond2}
\end{equation}

Since photoexcitation is fast for electrons to establish a hot-carrier distribution, we assume an effective electrononic temperature $T_e$ higher than the lattice temperature $T_{ph}$ for the pump-on case and equal to the lattce temperature $T_{ph}$ for the pump-off case. Photoconductivity is therefore the difference between the pump-on and pump-off conductivities $\displaystyle \Delta \sigma(\omega)= \sigma(T_e \neq T_{ph},\omega)-\sigma(T_e = T_{ph},\omega)$. By converting the integral over energy space to a summation over the $k$ space, we find:

\begin{equation}
\begin{split}
\frac{\Delta\sigma(\omega)}{G_0}=\frac{hV_F^2}{4N_kA}\Bigg\{
\sum_{\textbf{k}}\frac{-\frac{\partial f(\textbf{k},T_e \neq T_{ph})}{\partial \epsilon(\textbf{k})}}{\tau^{-1}(\textbf{k},T_e \neq T_{ph},T_{ph})-i\omega}\\
-\sum_{\textbf{k}}\frac{-\frac{\partial f(\textbf{k},T_e = T_{ph})}{\partial \epsilon(\textbf{k})}}{\tau^{-1}(\textbf{k},T_e = T_{ph},T_{ph})-i\omega}
\Bigg\}
\end{split}
\label{eqnphoto}
\end{equation}
where $G_0 = \frac{2e^2}{h}$ is the conductance quantum.

\begin{figure}[htb]
\includegraphics [width=\columnwidth]{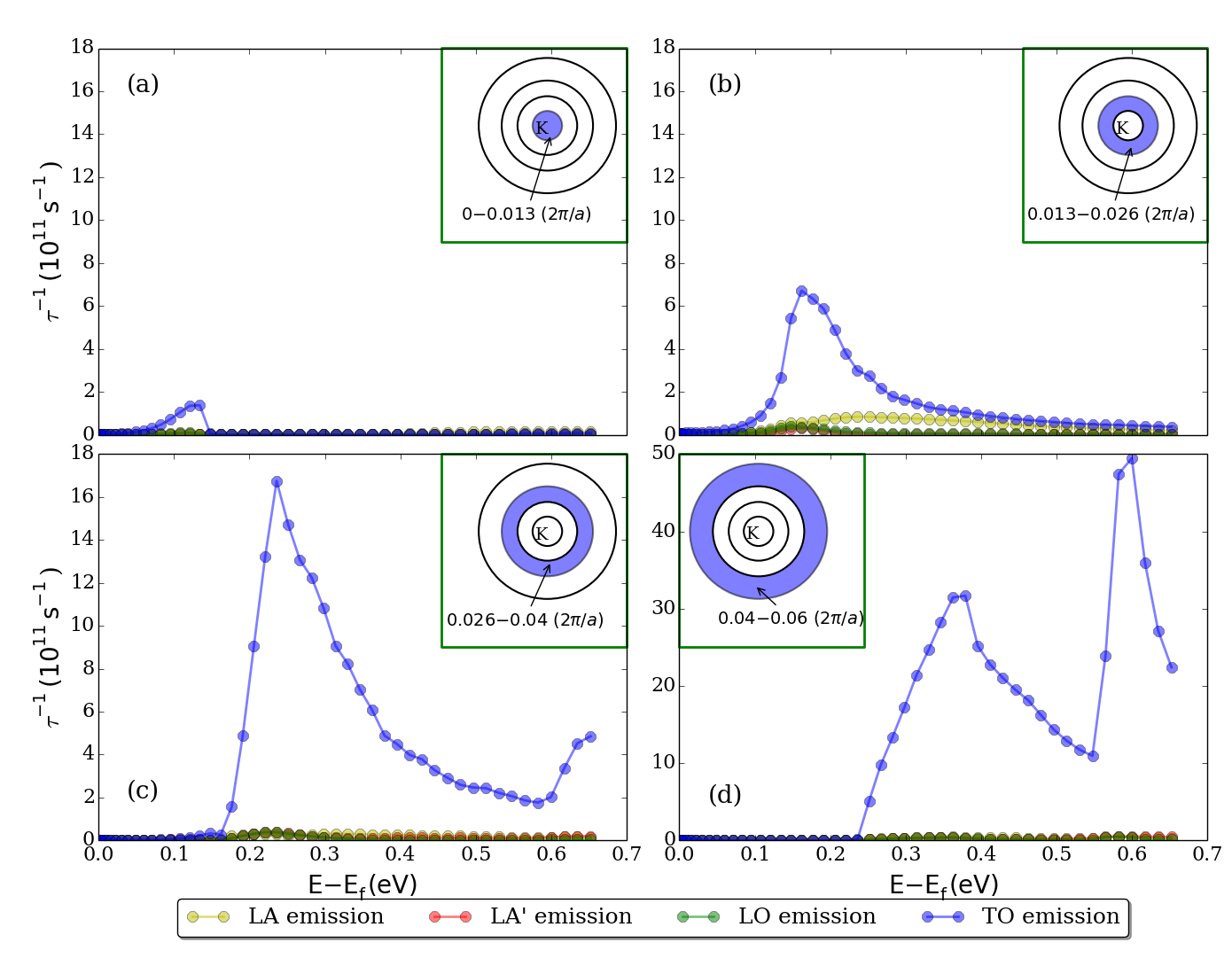}
\caption{Intrinsic bilayer graphene electron scattering rates at the phonon K point at $T_e$ = 300 K and $T_{ph}$ = 300 K. (a) Phonon wavevectors $\mathrm{0-0.013}\ 2\pi/a$ from the phonon K point. (b) Phonon wavectors $\mathrm{0.013-0.026}\ 2\pi/a$ from the phonon K point. (c) Phonon wavevectors $\mathrm{0.026-0.04}\ 2\pi/a$ from the phonon K point. (d) Phonon wavevectors $\mathrm{0.04-0.06}\ 2\pi/a$ from the phonon K point. Acoustic phonons rarely participate in scattering events at the phonon K point. Optical modes, especially the TO mode, dominate over all energy ranges.}

\label{fig:F3}
\end{figure}

\section{\label{sec:level4}IV. RESULTS AND DISCUSSION}

\begin{figure}[htb]
\includegraphics [width=\columnwidth]{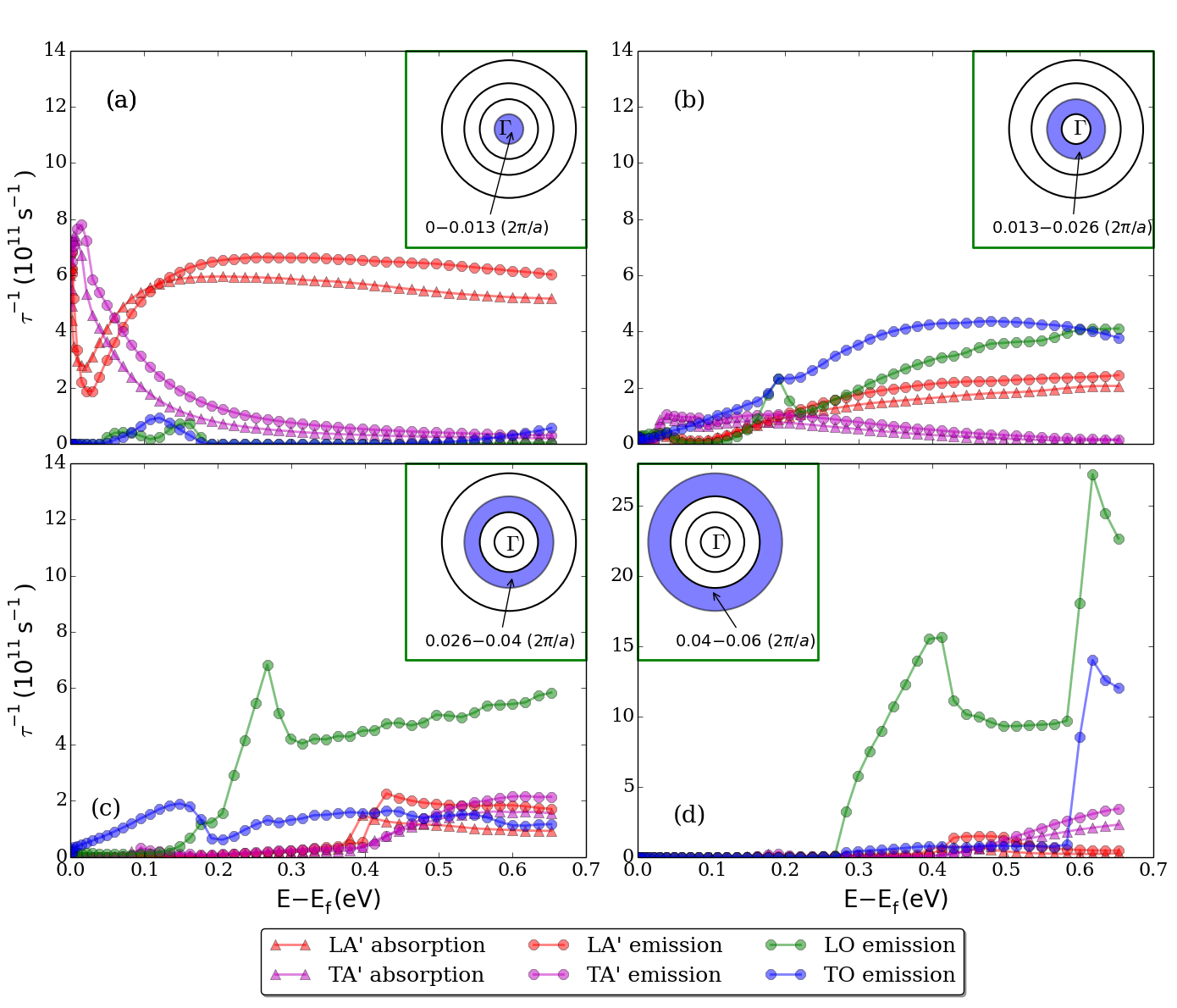}
\caption{Intrinsic bilayer graphene electron scattering rates near the phonon $\Gamma$ point at $T_e = 700$ K and $T_{ph}$ = 300 K. (a) Phonon wavevectors $\mathrm{0-0.013}\ 2\pi/a$. (b) Phonon wavectors $\mathrm{0.013-0.026}\ 2\pi/a$. (c) Phonon wavevectors $\mathrm{0.026-0.04}\ 2\pi/a$. (d) Phonon wavevectors $\mathrm{0.04-0.06}\ 2\pi/a$. The increase of electron temperature does not change acoustic mode scattering rates, but does cause more optical mode scattering in the low energy range.}

\label{fig:F4}
\end{figure}

In Bernal stacked bilayer graphene, half the atoms in the upper layer sit directly above the atoms in the lower layer while the other half lie at the centers of hexagons in the lower layer. The electronic structure was computed with the Quantum-Espresso package \cite{giannozzi2009quantum} using a norm-conserving pseudopotential in the local density approximation. A cutoff energy of 140 Ry and a Monkhorst-Pack $24\times 24\times 1$ k-space grid were chosen in the self-consistent calculations. The predicted interlayer distance is 3.3 \AA. Phonon dispersion calculations were performed with DFPT and a $12 \times 12 \times 1$ q-space grid. Due to the high computational cost of calculating energies and coupling matrix elements, we use an interpolation scheme based on maximally localized Wannier functions on a dense $1000\times 1000\times 1$ k-mesh and q-mesh with the EPW package \cite{ponce2016epw}. Our calculations for electronic band structure and phonon dispersion of bilayer graphene are shown in Fig.\ \ref{fig:F1}. Fig.\ \ref{fig:F1}(a) demonstrates that the Wannier-interpolated band structure completely overlaps the DFT calculated band structure within 1 eV of the Fermi level.

In single-layer graphene, the electron-phonon coupling matrix elements for TO and LO modes are significantly greater than those of acoustic modes in the low electron energy range, causing an order of magnitude larger scattering rates \cite{vallabhaneni2016reliability}. However, there are two major differences in the case of bilayer graphene. The first is that the electron-phonon coupling matrix elements are significantly smaller than single layer graphene, possibly due to the splitting of the two conduction bands and valence bands (see Fig.\ \ref{fig:F1}\ (a)). The second is the emergence of LA$^\prime $ and TA$^\prime$ modes, as indicated by the inset of Fig.\ \ref{fig:F1}\ (b). 

\begin{figure}[htb]
\includegraphics [width=\columnwidth]{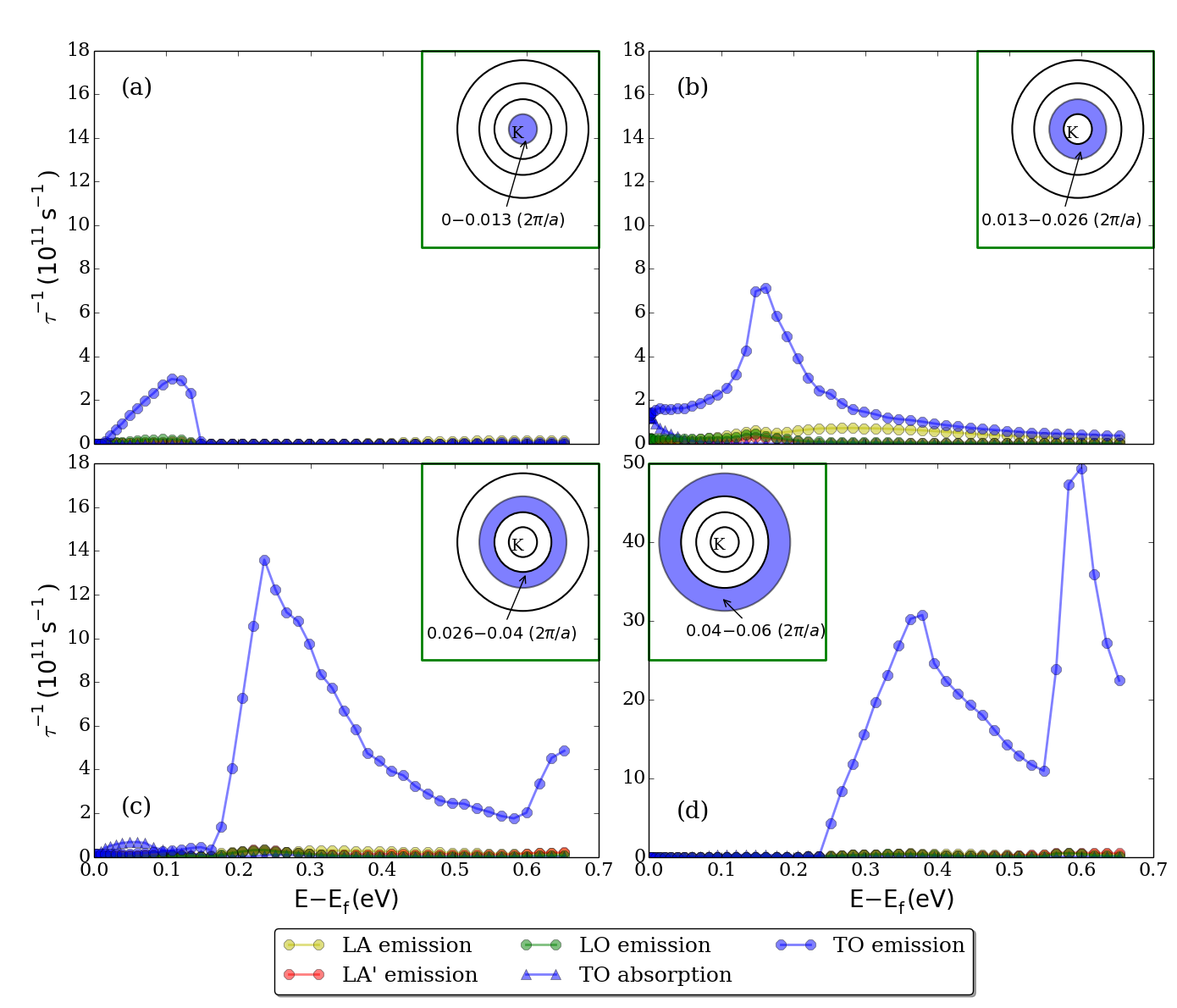}
\caption{Intrinsic bilayer graphene electron scattering rates at the phonon K point at $T_e$ = 700 K and $T_{ph}$ = 300 K. More TO mode scattering occurs due to the expansion of the Fermi window. (a) Phonon wavevectors $\mathrm{0-0.013}\ 2\pi/a$ from the phonon K point. (b) Phonon wavectors $\mathrm{0.013-0.026}\ 2\pi/a$ from the phonon K point. (c) Phonon wavevectors $\mathrm{0.026-0.04}\ 2\pi/a$ from the phonon K point. TO absorption emerges. (d) Phonon wavevectors $\mathrm{0.04-0.06}\ 2\pi/a$ from the phonon K point. }

\label{fig:F5}
\end{figure}

Because of the constraints of energy and momentum conservation, electrons mainly interact with $\Gamma$-point and K-point phonons. To differentiate between different phonon branches and to understand how phonons of different wavevectors couple with electrons, we chose a small region of radius $0.06\times 2\pi/a$ (a is the lattice constant) near the phonon $\Gamma$ and K points in reciprocal space and split it into four rings. The following calculations of scattering rates for a specific electronic state in the phase space are obtained by summing over all the phonon wavevectors within the corresponding rings on a dense 1000x1000x1 q-mesh. To highlight the effects of different phonon modes, the choices of the ring radii are based on the intersections of phonon branches near the phonon $\Gamma$ point and K point as indicated by Fig.\ \ref{fig:F1} (b), assuming isotropic dispersion.
The shaded blue regions in the insets of Figs.\ \ref{fig:F2}, \ref{fig:F3}, \ref{fig:F4}, and \ref{fig:F5} indicate the corresponding range of phonon wavevectors. The scattering rates were calculated for intrinsic bilayer graphene from first principles according to Eq. \ref{eqntau}. The subscript $m\textbf{k}$ is omitted because the scattering rates are plotted for electrons in the lower conduction band along the Dirac K - $\Gamma$ path. 

\begin{figure}[htb]
\includegraphics [width=\columnwidth]{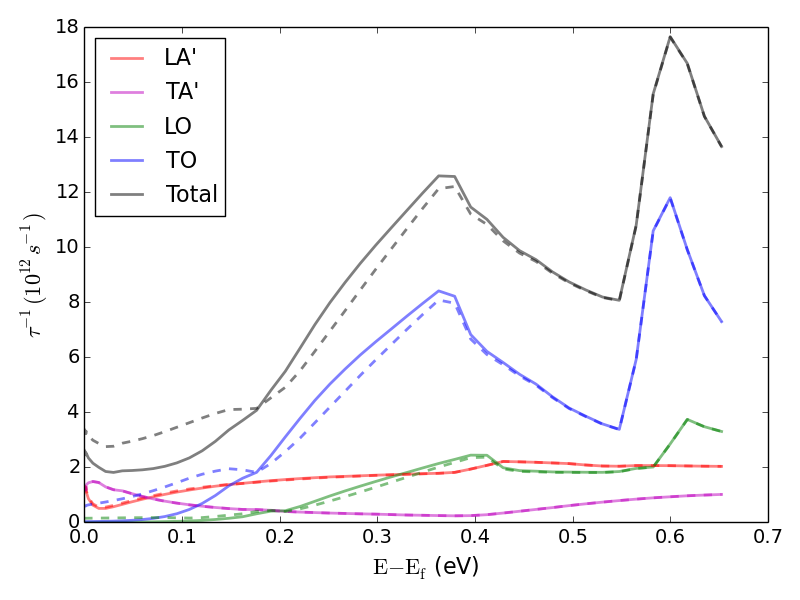}
\caption{Bilayer graphene electron-phonon scattering rates from phonon $\Gamma$ and K points at 300 K and 700 K. Solid lines denote scattering rates at $T_e$ = 300 K and dashed lines denote scattering rates at $T_e$ = 700 K. }

\label{fig:F10} 
\end{figure}

Fig.\ \ref{fig:F2}\ shows that in the low electron energy range ($<$ 200 meV) near the phonon $\Gamma$ point, LA$^\prime$ and TA$^\prime$ modes are dominant and confined to small phonon wavevectors ($<0.026 \times 2\pi/a$). Unlike single-layer graphene, bilayer graphene scattering rates are not linearly proportional to the energies from the Dirac \textbf{K} point because LA$^\prime$ and TA$^\prime $ modes do not exhibit a linear dispersion near the phonon $\Gamma$ point(see Fig.\ \ref{fig:F1}\ (b)).

In Fig.\ \ref{fig:F2}\ (a), the absorption process is slightly stronger than the emission process at $E-E_f <100$ meV for the LA$^\prime$ mode because final states have higher densities of states as shown in the inset of Fig.\ \ref{fig:F1} (a). However, for high energy electrons the emission process is more active because the final states have higher occupations. This phenomenon is also observed in Fig.\ \ref{fig:F4}, where the transition happens at higher electron energy. As shown in Fig.\ \ref{fig:F2}\ (b), (c) and (d), optical modes only participate in the emission process, because almost no phonons with energies higher than 200 meV are excited at 300 K. 

Fig.\ \ref{fig:F3} shows scattering rates due to interactions with K-point phonons at $T_e = 300$ K and $T_{ph} = 300$ K.  Similar to results at phonon $\Gamma$ point, significant LO and TO mode absorption does not occur. Electrons primarily interact with TO mode phonons at the K point due to large electron-phonon coupling strength. 

We also investigate temperature effects on scattering rates by increasing the electron temperature $T_e$ from 300 K to 700 K with the phonon temperature $T_{ph}$ fixed at 300 K in Figs.\ \ref{fig:F4} and \ref{fig:F5}. Fig.\ \ref{fig:F4} demonstrates that increasing the electron temperature does not influence the acoustic modes while causing more optical phonon emission at the phonon $\Gamma$ point in the low energy range as a consequence of more available final states, but the increase compared with Fig.\ \ref{fig:F2} is not very significant. The effects on K-point phonons are similar, as shown in Fig.\ \ref{fig:F5}. The scattering starts to involve the TO absorption process due to the expansion of the Fermi window. The total scattering rates from phonon $\Gamma$ and K points are shown in Fig.\ \ref{fig:F10}. At low electron energies, the scattering rates increase with electron temperature, and the major contribution is from increased TO phonon scattering.

\begin{figure}[htb]
\includegraphics [width=\columnwidth]{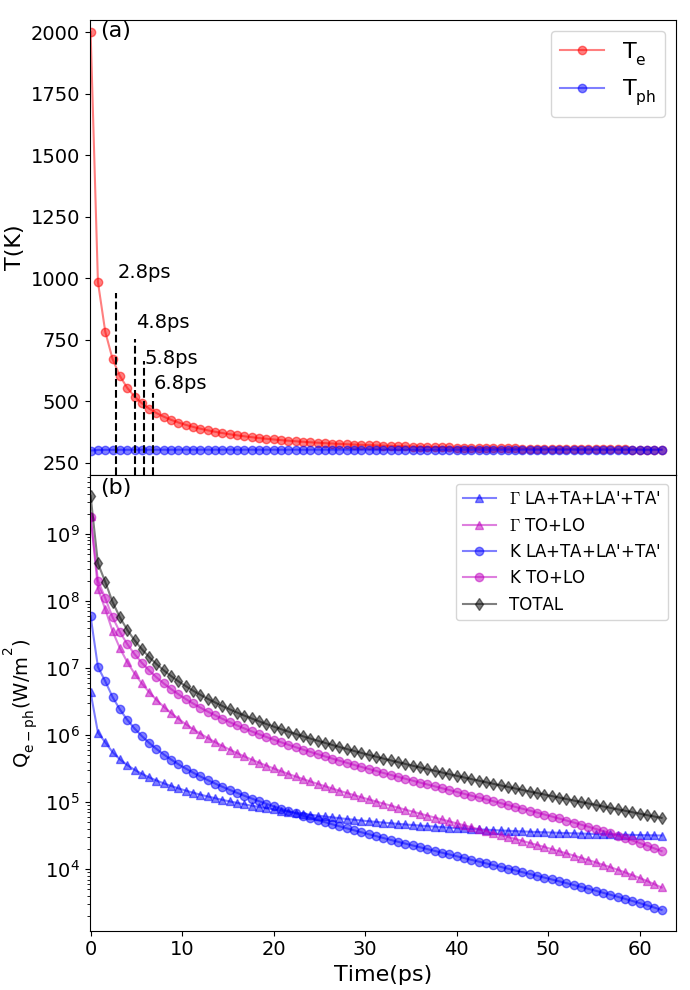}
\caption{Two-temperature model for bilayer graphene with $E_{fermi}$ = 10 meV. (a) Electron and lattice temperature changes as functions of time. The thermalization process takes 60 ps to reach equilibrium. The electron temperature drops drastically (by approx. 800 K) in the first 0.4 ps while the lattice temperature remains almost constant around 303 K. (b) Heat flux to the phonon $\Gamma$ point and $\mathrm{K}$ point acoustic and optical modes. Most heat is diverted to K-point phonon optical modes. After 42 ps, heat transfer to $\Gamma$ point acoustic modes exceeds that to $\Gamma$ point optical modes.}

\label{fig:F6} 
\end{figure}

To explore the effects of doping levels on photoconductivity, we first calculate the temperature change with time and the corresponding heat flow into $\Gamma$-point and K-point acoustic and optical phonons based on Eqs. \ref{eqnenergy} and \ref{eqnevol}. We choose a starting electron temperature $T_e = 2000$ K and lattice temperature $T_{ph} = 300$ K, and allow the system to relax for bilayer graphene with Fermi levels 10 meV, 20 meV, 40 meV, and 60 meV respectively. Fig.\ \ref{fig:F6} (a) shows that the thermalization process for bilayer graphene with $E_{fermi}=10$ meV requires 60 ps to reach equilibrium, and the phonon temperature remains almost constant around 303 K because of the large phonon heat capacity. The electron temperature drops rapidly to 1200 K during the first 0.4 ps. Therefore even if the initial electron temperature may not be a precise estimate, further calculations are not severely affected. From Fig.\ \ref{fig:F6} (b), at temperatures above room temperature, most energy loss by electrons flows to optical phonons near the phonon K point because of strong interactions with electrons indicated by Figs.\ \ref{fig:F3} and \ref{fig:F5} and because the population of these modes requires high energies. As the electron temperature approaches 303 K after around 42 ps, the energy diverted to $\Gamma$ point optical modes is lower than acoustic modes because optical modes rarely interact with low energy electrons near the $\Gamma$ point. 
\begin{figure}[htb]
\includegraphics [width=\columnwidth]{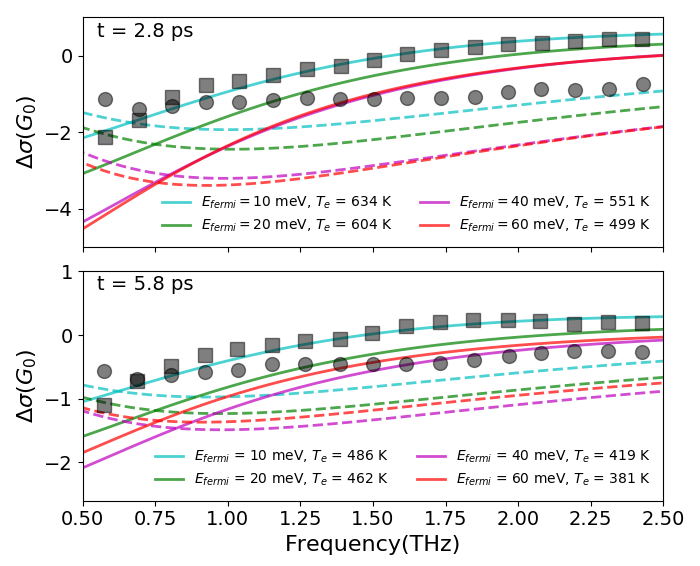}
\caption{Photoconductivity evolution. Squares and circles are experimental data \cite{kar2018ultrafast} for real and imaginary parts of photoconductivity, solid lines denote calculated real parts,  and dashed lines denote calculated imaginary parts. The electron and lattice temperatures at each time step are obtained from our previous two-temperature calculations.}
\label{fig:F7}
\end{figure}

\begin{figure}[htb]
\includegraphics [width=\columnwidth]{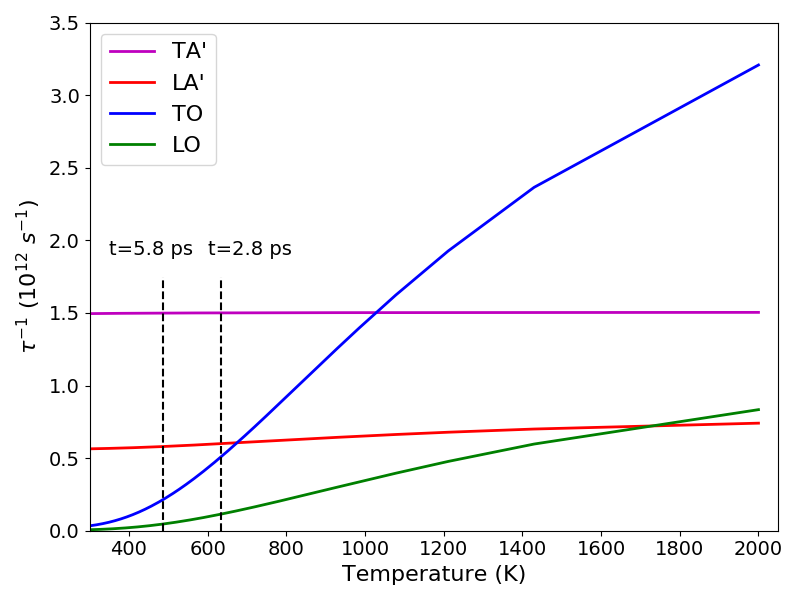}
\caption{Electron scattering rates at Fermi level in bilayer graphene with $E_{fermi}$ = 10 meV due to interactions with  different phonon branches . Dashed lines denote scattering rates at $t$ = 2.8 ps ($T_e$ = 634 K) and $t$ = 5.8 ps ($T_e$ = 486 K). }

\label{fig:F8}
\end{figure}

Using the electron and phonon temperatures obtained from the two-temperature model, we calculated photoconductivities at four specific times, $t$ = 2.8 ps, $t$= 4.8 ps, $t$ = 5.8 ps and $t$ = 6.8 ps, and compared with prior experimental results. We plot only $t$ = 2.8 ps and $t$ = 5.8 ps in this paper as shown in Fig.\ \ref{fig:F7}.  The case $E_{fermi}$ = 10 meV matches with experimental values best in terms of both trend and intersection of the real and imaginary parts at all four times. The negative imaginary part can be qualitatively explained by an increase of temperature that amplifies overall scattering rates. The imaginary part goes to 0 as the probe frequency increases to values considerably larger than the overall scattering rates, because the effect of increased scattering become insubstantial and eventually diminishes. At $t$ = 2.8 ps,  the real part of photoconductivity decreases significantly with doping level at low probe frequencies due to enhanced electron-phonon scattering.  Note that at $t$ = 5.8 ps, the conductivity for $E_{fermi}$ = 40 meV is higher than that for $E_{fermi}$ = 60 meV because the heat loss to phonons is more severe for $E_{fermi}$ = 60 meV. Thus the electron temperature is lower than that for $E_{fermi}$ = 40 meV at the same time even though the former has higher density of states. As the relaxation proceeds, the real part of photoconductivity increases because the electron temperature decreases and weakens electron-phonon scattering. Fig.\ \ref{fig:F8} shows that from $t$ = 2.8 ps to $t$ = 5.8 ps, electron-phonon scattering rates decrease and the major reason for this decrease is the reduction of TO phonon scattering. Scattering from acoustic phonons is insensitive to temperature but that from optical modes depends strongly on temperature.  At temperatures higher than 1000 K, TO mode phonon scattering dominates over other branches. At temperatures lower than 600 K, TA$^\prime$ and LA$^\prime$ mode scattering dominates over TO and LO mode phonons which does not contradict the result shown in Fig.\ \ref{fig:F6} . Heat flux to acoustic phonons is significantly smaller than that to optical phonons from $t$ = 2.8 ps to $t$ = 6.8 ps because optical phonons require more energy to be populated. The real part also increases with probe frequency because the scattering time is comparably longer for carriers to react to electric field oscillations. Similarly, differences in conductivity between different doping levels are smaller at high frequencies where the scattering effects are less prominent.

Assuming $T_e$ = $T_L$ = 300 K and following the approach in \cite{kaasbjerg2012unraveling}, we combine contributions from the deformation potential and gauge field, and then extract an effective deformation potential coupled to a single phonon mode. The extracted $E_{1, \mathrm{eff}}$ is $22$ eV calculated with Eq.\ \ref{redavg} for an effective phonon group velocity of $v_{ph} = 2.0\times 10^4$ m/s. Fig.\ \ref{fig:F2} indicates that both LO and TO modes contribute to scattering, and we use a single phonon optical deformation potential to represent LO and TO modes at the phonon $\Gamma$ point. The corresponding phonon frequency $\omega_{\mathrm{o, LO/TO},\Gamma}$ is 200 meV. At the phonon K point, electrons mainly interact with the TO mode; therefore we use the K-point TO phonon frequency $\omega_{\mathrm{o, TO, K}} = 166$ meV in Eq.\ \ref{tauopt}. The calculated $D_{\mathrm{o,LO/TO,\Gamma}}$ is 1.8 $\mathrm{eV/\AA}$, and $D_{\mathrm{o,TO,K}}$ is 2.5  $\mathrm{eV/\AA}$.  

\begin{figure}[htb]
\includegraphics [width=\columnwidth]{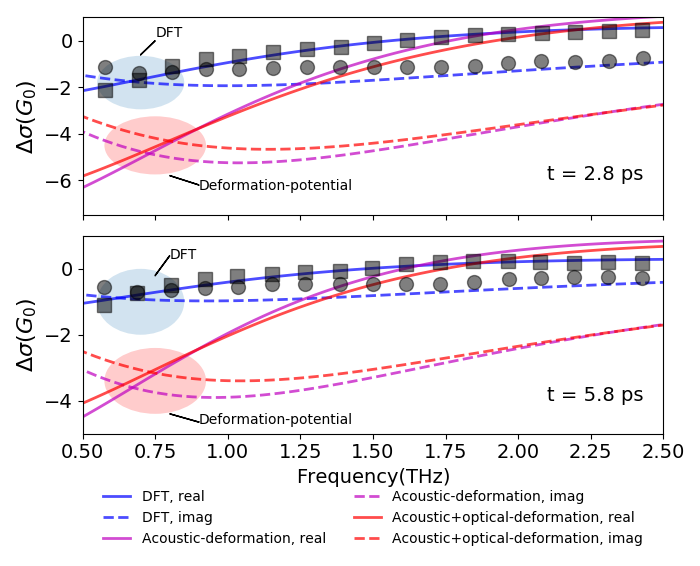}
\caption{Comparison of photoconductivity calculations from DFT and the deformation-potential method. Squares and circles are experimental data \cite{kar2018ultrafast} for real and imaginary parts of photoconductivity. Solid lines and dashed lines denote calculated real and imaginary components of photoconductivity, respectively. The conductivity predicted by the deformation-potential model deviates from the experimental data significantly in the low probe frequency regime.}

\label{fig:F9}
\end{figure}

We further calculated deformation-potential scattering rates according to Eq. \ref{taufinal} with the derived deformation potentials and calculated photoconductivity for $E_{fermi}$ = 10 meV at four different times. As shown in Fig.\ \ref{fig:F9}, the real part of photoconductivity calculated using the acoustic-deformation potential deviates significantly from the experimental data, as opposed to DFT predictions. The inclusion of an optical-deformation potential slightly improves the predictions for both real and imaginary parts; however the deviation from experiments and DFT results is still large. This indicates that momentum-dependent electron-phonon scattering potential and scattering rates are likely required to achieve accuracy in photoconductivity calculations. 

Because defects indeed exist in the experimental bilayer graphene samples even though the concentration may be low, we expect a discrepancy between our calculations and measurements. Apart from electron-phonon scattering, short-range scattering also occurs in reality. Another factor is the indeterminism of the electron temperature, without which the calculated photoconductivity is not accurate. The two-temperature model could also be a source of discrepancy, but its effects are likely not significant. Phonon temperatures could depend on branches and their positions in reciprocal space; however they should always be around 300 K because of their large heat capacities.

\section{\label{sec:level5} V. CONCLUSIONS}
This paper demonstrates first-principles methods for the calculation of eletron-phonon coupling, based on which an effective acoustic deformation potential $E_{1, \mathrm{eff}} = 22$ eV is estimated for bilayer graphene. The Drude model with DFT-calculated scattering rates predicts the correct trend for real and imaginary parts of photoconductivity. By increasing the doping level, the electron-phonon scattering is enhanced especially for low probe frequencies.  The comparison between our DFT and deformation-potential approach calculations indicates that first-principles methods result in less deviations from experiments.  The small deviations from experimental values could derive from several causes such as defects in the sample, short-range scattering, and electron temperature inaccuracies. The initial electron temperature is related to the incident light frequencies, intensities and the sample area illuminated. A possible direction of the future work could be determining the initial electron temperature accounting for fast electron-electron scattering that involves many-body effects. 

\section{\label{sec:acknow} ACKNOWLEDGMENTS}
Y. G. thanks Dr. Yanguang Zhou for fruitful discussions and comments. This work used the Extreme Science and Engineering Discovery Environment (XSEDE), which is supported by National Science Foundation grant number ACI-1548562 and computational and storage services associated with the Hoffman2 Shared Cluster provided by UCLA Institute for Digital Research and Education’s Research Technology Group.

\bibliographystyle{apsrev4-1}

\bibliography{bib1}






%
\end{document}